\documentstyle[12pt]{article}
\def\tilde{\widetilde}
\def\ta{U^{1}}
\def\tb{U^{2}}
\def\ga{G^{1}_{+}(n)}
\def\gb{G^{2}_{+}(m)}
\def\gaa{G^{1}_{-}(n)}
\def\gbb{G^{2}_{-}(m)}
\def\La{L^{1}_{+}}
\def\Lb{L^{2}_{+}}
\def\laa{L^{1}_{-}}
\def\lbb{L^{2}_{-}}
\def\Lc{L^{1}_{\pm}}

\def\mc{\tilde L^{1}_{\pm}}
\def\r{r_{+}}
\def\ra{r_{-}}
\def\rb{r_{\pm}}
\def\p{\partial}

\def\tr{\,{\rm tr}\,}

\def\a{\alpha}
\def\b{\beta}
\def\am{A_{\mu}}

\def\an{A_{\nu}}

\def\G{Gauss-law constraints }
\def\ui{U({\bf n};i)}

\def\ei{E({\bf n};i)}

\def\n{{\bf n}}
\def\m{{\bf m}}

\def\la{\lambda^{a}}
\def\lb{\lambda^{b}}
\def\lc{\lambda^{c}}

\def\be{\begin{equation}}
\def\ee{\end{equation}}
\def\bea{\begin{eqnarray}}
\def\eea{\end{eqnarray}}
\def\Y{Yang-Mills }

\def\a{\alpha}
\def\h{the Heisenberg double }
\def\g{\gamma}
\vspace{5cm}

\title{  \hfill{LMU-TPW 95-1}
\\Gauge-invariant Hamiltonian formulation of lattice Yang-Mills theory and the
Heisenberg double}
\vspace{.7cm}
\author{ \mbox{}
\\ S.A.Frolov\thanks{Alexander von Humboldt fellow}
\mbox{} \\ \vspace{0.4cm} Section Physik, Munich University
\vspace{-0.5cm} \mbox{} \\ Theresienstr.37, 80333 Munich, Germany
\thanks{Permanent address:\ Steklov Mathematical Institute, Vavilov st.42,
GSP-1, 117966 Moscow, RUSSIA}
\mbox{} \\
\date{}}

\begin{document}
\maketitle
\vspace{4.5cm}
\begin{abstract}
It it known that to get the usual Hamiltonian formulation of lattice \Y theory
in the temporal gauge $A_{0}=0$ one should place on
every link the cotangent bundle of a Lie group. The cotangent bundle may be
considered as a limiting case of a so called Heisenberg double
of a Lie group which is one of the basic objects in the theory of Lie-Poisson
and quantum groups. It is shown in the paper that there is a generalization
of the usual Hamiltonian formulation to the case of \h .
\end{abstract}

 \section{Introduction}

Lattice regularization of gauge theories is at present the only known
nonperturbative regularization. There are two  possible ways of discretization
of the theories.
In the approach of Wilson \cite{w} one consideres Euclidean formulation of
gauge
theories and discretizes all space-time, thus replacing the theory by some
statistical mechanics model. In the approach of Kogut and Susskind \cite{ks}
one
consideres gauge theories in the Minkowskian space-time and in the first-order
(Hamiltonian) formulation and discretizes only the space directions remaining
the time variable continuous. In this case the \Y theory is replaced by some
classical mechanics model with constraints.
Passing from the continuous \Y theory to a lattice model one replaces the
infinite dimensional phase space of the \Y theory by the finite dimensional
phase space of the lattice model. Thus although the first approach is more
suitable for practical calculations, the second one seems to be more
appropriate
for the study of different nonequivalent representations of the infinite
dimensional Heisenberg algebra of the canonical variables of the \Y theory.

In their paper \cite{ks} Kogut and Susskind considered from the very beginning
quantum theory of lattice \Y model in the temporal gauge $A_{0}=0$. However it
appeares to be useful to begin with the gauge-invariant formulation of the
classical lattice  \Y theory and then to quantize it in some gauge (for example
in the temporal gauge). In the present paper we consider such a formulation of
the \Y theory and as an example of a useful application present a very simple
solution of the (1+1)-dimensional \Y theory on a cylinder. To get this
formulation, which seems to be known, one should place on every link the
cotangent bundle of a Lie group. Thus the phase space of this lattice gauge
theory is the direct product of the cotangent bundles over all links. The
cotangent bundle can be considered as a limiting case of a so called Heisenberg
double $D_{+}^{\g}$
of a Lie group which is one of the basic objects in the theory of Lie-Poisson
and quantum groups \cite{d1,s1,d2,frt,s2}. In the last section of this paper we
show that there is a generalization of the usual Hamiltonian formulation of
lattice gauge theory to the case of \h. We consider the (1+1)-  and
(2+1)-dimensional $SL(N,C)$ \Y theory and show that one can single out the real
forms $SU(N)$ and $SL(N,R)$ by imposing the corresponding involutions of \h.
The
lattice theory obtained has two parameters $a$ and $\g$ and coincides with the
continuous \Y theory in the continuum limit $a \to 0$, $\g/a \to const$. At the
end of the paper we make some remarks on quantization of this lattice gauge
theory.

\section{Lattice Yang-Mills theory and cotangent bundle}

In this section we remind some known results concerning the Hamiltonian
formulation of the lattice (d+1)-dimensional \Y theory described by the
following action
\be
S=\frac {1}{8e^{2}}\int dx\,\tr F_{\mu\nu}^{2}
\label{1}
\ee
Here $F_{\mu\nu}=\p_{\mu}\an-\p_{\nu}\am + [\am,\an]$ and $\am=\am^{a}\la$ is a
gauge field with values in the Lie algebra $g$ of a simple Lie group $G$,
$\la$ are generators of the Lie algebra with the following relations
\be
\tr \la \lb = -2 \eta^{ab} , \quad [\la ,\lb ]=f^{abc} \lc
\label{2}
\ee
where $\eta^{ab}$ is the Killing tensor of the group $G$.

\noindent The action (\ref{1}) can be rewritten in the first-order formulation
as follows
\be
S=-\frac {1}{2}\int dtd{\bf x}\,\tr ( E_{i}\frac {\p A_{i}}{\p t} - \frac
{e^{2}}{2}E_{i}^{2} - \frac {1}{4e^{2}}F_{ij}^{2} + A_{0}(\p _{i} E_{i} +
[A_{i}, E_{i}])),
\label{3}
\ee
where $i,j=1,...,d$ are space indices.

\noindent It is obvious from eq.(\ref{3}) that $E_{i}$ and $A_{i}$ are
canonically-conjugated momenta and coordinates with the standard Poisson
structure
\be
\{A_{i}^{a}({\bf x}), E_{j}^{b}({\bf y})\}=\eta_{ab}\delta_{ij}\delta({\bf
x}-{\bf y}),
\label{4}
\ee
\be
H=-\frac {1}{2}\int d{\bf x}\,\tr (\frac {1}{2}E_{i}^{2} +\frac
{1}{4}F_{ij}^{2})
\label{5}
\ee
is the Hamiltonian of the system, $A_{0}$ is a Lagrange multiplier and
\be
G=\p_{i} E_{i} + [A_{i}, E_{i}]=G^{a}\la
\label{6}
\ee
are the \G forming the gauge algebra
\be
\{G^{a}({\bf x}), G^{b}({\bf y})\}=f^{abc}G^{c}({\bf x})\delta({\bf x}-{\bf
y}),
\label{7}
\ee
Thus the classical \Y action describes a system with first-class constraints
and
an infinitely-dimensional phase space which can be presented as
$\prod_{{\bf x}} (g\times g)_{{\bf x}}=\prod_{{\bf x}} (T^{*}g)_{{\bf x}}$.

Eqs.(\ref{4}) and (\ref{7}) can be rewritten in the following form which will
be
used in the paper
\be
\{E_{i}^{1}({\bf x}), A_{j}^{2}({\bf y})\}=C\delta_{ij}\delta({\bf x}-{\bf y}),
\label{4a}
\ee
\be
\{G^{1}({\bf x}), G^{2}({\bf y})\}=\frac {1}{2}[G^{1}({\bf x}) - G^{2}({\bf
y}),C]\delta({\bf x}-{\bf y}),
\label{7a}
\ee
In eqs.(\ref{4a}) and (\ref{7a}) we use the standard notions from the theory of
quantum groups \cite{frt}: for any matrix $A$ acting in some space $V$ one can
construct two matrices $A^{1}=A\otimes id $ and $A^{2}=id\otimes A $ acting in
the space $V\otimes V$, and matrix $C$ is the tensor Casimir operator
of the Lie algebra: $C=-\eta_{ab}\la\otimes\lb$. For the case of the $SL(2)$
algebra $C=\sum_{a=1}^{3} \sigma^{a}\otimes\sigma^{a}$, where $\sigma^{a}$ are
Pauli matrices.

Let us now introduce the regular hyper-cubic space lattice. An arbitrary vertex
of the lattice is denoted by a vector ${\bf n}=(n_{1},...,n_{d})$ with
integers $n_{i}$, the orthonormal lattice vectors are denoted ${\bf e}_{i}$,
$i=1,...,d$ and an arbitrary link is denoted by a vector ${\bf n}$ and a
lattice
vector ${\bf e}_{i}$: $({\bf n},{\bf e}_{i})$ or $({\bf n},i)$.
To get a gauge-invariant lattice \Y theory, we place on each vertex an
algebra-valued Lagrange multiplier $A_{0}({\bf n})$
 and on each link $({\bf n},{\bf e}_{i})$
a group-valued field $\ui$ and an algebra-valued field $\ei$. Then the lattice
\Y action which is invariant with respect to the following gauge
transformations
\bea
\ui &\to & g^{-1}(\n)\ui g(\n + {\bf e}_{i}) \nonumber\\
\ei &\to & g^{-1}(\n)\ei g(\n) \nonumber\\
A_{0}({\bf n}) &\to & g^{-1}(\n)A_{0}({\bf n})g(\n) + g^{-1}(\n)\frac {dg(\n)}
{dt},
\label{8}
\eea
can be written as follows
\bea
S_{lat}=&-&\frac {1}{2}\tr\big[ \sum_{\n,i} \big( \ei \frac {d\ui}{dt}
U^{-1}(\n
,i) - \frac {e^{2}}{2} a^{2-d}E^{2}(\n ,i) \big) \nonumber \\
&+& \sum_{\n}A_{0}({\bf n} )\sum_{i} \big( \ei - U^{-1}(\n -{\bf e}_{i};i)E(\n
-{\bf e}_{i};i)U(\n -{\bf e}_{i};i)\big) \big] \nonumber\\
&-&\frac  {a^{d-4}}{2e^{2}} \sum_{plaqettes}\big(W(\Box) + W^{*}(\Box)\big)
\label{9}
\eea
Here $W(\Box) $ is the usual Wilson term
\be
W(\Box)= \tr \ui U(\n +{\bf e}_{i};j)U^{-1}(\n +{\bf e}_{j};i)U^{-1}(\n ;j)
\label{10}
\ee
In the continuum limit
\bea
a \to  0, \qquad  A_{0}({\bf n}) &\to & A_{0}(a{\bf n}) \nonumber\\
\ei &\to &a^{d-1}E_{i}(a\n)     \nonumber\\
 \ui &\to &1+ aA_{i}(a\n),
\label{11}
\eea
the action (\ref{9}) reduces to the action (\ref{3}).

The kinetic term in eq.(\ref{9}) defines the Poisson structure for the fields
$\ui$ and $\ei$
\bea
&& \{ U^{1}(\n;i),U^{2}(\m;j)\} =0 \nonumber\\
&& \{ E^{1}(\n;i),E^{2}(\m;j)\} =  \frac {1}{2}
[E^{1}(\n;i)-E^{2}(\m;j),C]\delta_{ij}\delta_{\n\m} \nonumber\\
&& \{ E^{1}(\n;i),U^{2}(\m;j)\} = CU^{2}(\m;j)\delta_{ij}\delta_{\n\m}
\label{12}
\eea
This Poisson structure being ultralocal coincides on every link with the
canonical Poisson structure of the cotangent bundle of the group $G$: $T^{*}G$.
Thus the
phase space of the regularized model is the direct product of $T^{*}G$ over all
links: $\prod_{(\n, {\bf e}_{i})} (T^{*}G)(\n,i)$.

Using eqs.(\ref{12}) one can easily calculate the Poisson bracket of the \G
\bea
&&G(\n)=\sum_{i} \big( \ei - U^{-1}(\n -{\bf e}_{i};i)E(\n -{\bf e}_{i};i)U(\n
-{\bf e}_{i};i)\big) \nonumber\\
&& \{ G^{1}(\n),G^{2}(\m)\} =  \frac {1}{2}
[G^{1}(\n)-G^{2}(\m),C]\delta_{\n\m}
\label{13}
\eea

One can see from eqs.(\ref{12})  and (\ref{13}) that the field $\ei$ should be
identified with the right-invariant momentum generating left gauge
transformations of the field
$\ui$ and the field $\tilde E(\n;i)=U^{-1}(\n;i)\ei\ui$ should be identified
with the left-invariant momentum generating right gauge transformations of the
field
$\ui$. It is not difficult to check that $\ei$ and $\tilde E(\n;i)$ have the
vanishing Poisson bracket. Let us note that in the continuum limit (\ref{11})
the
Poisson structure (\ref{12}) reduces to eq.(\ref{4a}). Now imposing the
temporal
gauge $A_{0}=0$ and quantizing the Poisson structure (\ref{12}) one gets the
model considered by
Kogut and Suskind \cite{ks}.

As an application of this gauge-invariant formulation let us consider
quantization of the (1+1)-dimensional \Y theory on a cylinder. In this case the
lattice action (\ref{9})
can be rewritten as follows
\bea
S_{lat}=&-&\frac {1}{2}\tr\big[ \sum_{n=1}^{N} \big( E(n)\frac {dU(n)}{dt}
U^{-1}(n) - \frac {e^{2}}{2} aE^{2}(n) \nonumber \\
&+&A_{0}(n)( E(n) - U^{-1}(n -1)E(n -1)U(n -1)\big)
 \big]
\label{14}
\eea
Due to the gauge invariance of the action one can impose the gauge condition
\be
U(n)=1,\quad n\not= N;\quad U(N)\equiv U
\label{15}
\ee
Then one can easily solve the \G $G(n)=0$ for $n\geq 2$ and get the following
solution
\be
 E(n)=E\qquad for \quad any \quad n
\label{a}
\ee
Inserting this solution to the first constraint $G(1)=E(1)-U^{-1}E(N)U$ one
gets
a residual constraint $G=E-U^{-1}EU$ which generates the residual gauge
transformations
\be
U \to g^{-1}Ug,\quad E \to g^{-1}Eg
\label{16}
\ee
Finally the action (\ref{14}) takes the form
\be
S_{lat}=-\frac {1}{2}\tr\big[ E\frac {dU}{dt}U^{-1} - e^{2}\pi RE^{2} +A_{0}( E
-U^{-1}EU) \big]
\label{17}
\ee
where $R=aN/2\pi$ is the radius of the circle.

It is worthwhile to note that the resulting action (\ref{17}), firstly obtained
in ref.\cite{r} by a different method, does not depend on the lattice length
$a$
and
thus is the exact action describing partially-reduced \Y theory.

This gauge-invariant Hamiltonian formulation of the lattice \Y theory was based
on the cotangent bundle of the group G. One could ask oneself whether it is
possible
to put on each link another phase space and on each vertex another first-class
\G and to get another lattice theory which in the continuum limit reduces to
the
continuous one. In the next section we show that such a possibility does exist
and is based on a phase space which is called the Heisenberg double of a Lie
group
and is one of the basic objects in the theory of Poisson-Lie and quantum
groups.

\section{Heisenberg double and lattice Yang-Mills theory}

In this section we present some simple results on the theory of \h which will
be
used later. More detailed discussion of the subject can be found in
refs.\cite{s1,s2,am} .

Let $G$ be a matrix algebraic group and $D=G\times G$. For definiteness we
consider the case of the $SL(N)$ group. Almost all elements $(x,y) \in D$ can
be
presented as follows

\be
(x,y)=(U,U)^{-1}(L_{+},L_{-}),
\label{2.1}
\ee
where $U \in G$, the matrices $L_{+}$ and $L_{-}$ are upper- and
lower-triangular, their diagonal parts $l_{+}$ and $l_{-}$ being inverse to
each
other: $l_{+}l_{-}=1$.

Let all of the matrices be in the exact matrix representation $\rho$, $V$ of
the
group $G$. Then the algebra of functions on the group $D$ is generated by the
matrix elements $x_{ij}$ and $y_{ij}$. The matrices $L_{\pm}$ and $U$ can be
considered as almost everywhere regular functions of $x$ and $y$. Therefore,
the
matrix elements $L_{\pm ij}$ and
$U_{ij}$ define another system of generators of the algebra $Fun D$. We define
the Poisson structure on the group $D$ in terms of the generators $L_{\pm}$ and
$U$ as follows \cite{s2}
\be
\{\ta ,\tb \} =\g [\rb ,\ta\tb ]
\label{2.2a}
\ee
\bea
&&\{\La ,\Lb \} =\g [\rb ,\La\Lb ] \nonumber\\
&&\{\laa ,\lbb \} =\g [\rb ,\laa\lbb] \nonumber\\
&&\{\La ,\lbb \} =\g [\r ,\La\lbb]
\label{2.2b}
\eea
\bea
&&\{\La ,\tb \} =\g \r \La\tb \nonumber\\
&&\{\laa ,\tb\} =\g \ra \laa\tb
\label{2.2c}
\eea
Here $\g$ is an arbitrary complex parameter, $\rb$ are classical $r$-matrices
which satisfy the classical Yang-Baxter equation and the following relations
\be
\ra =-P\r P
\label{2.3a}
\ee
\be
\r -\ra =C
\label{2.3b}
\ee
where $P$ is a permutation in the tensor product $V\otimes V$ ($Pa\otimes
b=b\otimes a$).
For example in the case of the $SL(2)$ group
\be
\r =\left( \begin{array}{cccc} \frac {1}{2} & 0 &  0 & 0 \\
0 & - \frac {1}{2} & 2 & 0 \\
0 & 0 &  - \frac {1}{2} & 0 \\
0 & 0 & 0 & \frac {1}{2} \end{array} \right)
\label{2.4}
\ee
The group $D$ endowed with the Poisson structure (\ref{2.2a}-\ref{2.2c}) is
called the Heisenberg double $D_{+}^{\g}$ of the group $G$. To understand the
relation of \h $D_{+}^{\g}$ to the cotangent bundle of the group $G$ it is
convinient to use the matrix $L=L_{+}L_{-}^{-1}$ \cite{sr}. The Poisson
brackets
for $L$ and $U$ can be written as follows
\bea
\frac {1}{\g} \{L^{1},L^{2}\} =L^{1}L^{2}\ra +\r L^{1}L^{2} -L^{1}\ra L^{2} -
L^{2}\r L^{1} \nonumber \\
\frac {1}{\g} \{U^{1},L^{2}\} =\ra U^{1}L^{2} - L^{2}\r U^{1}
\label{2.5}
\eea
If one consideres now the limit $\g \to 0$ and $L \to 1+\g E$, $L_{\pm} \to
1+\g
E_{\pm}$ one gets the canonical Poisson structure of the cotangent bundle
$T^{*}G$ (see eq.(2.14)). Thus one can see that $L$ is an analog of the
right-invariant momentum and one can introduce the left-invariant momentum
$\tilde L$ by the following equation
\be
\tilde L =U^{-1}LU
\label{2.6}
\ee
The matrix $\tilde L$ can be decomposed into the product of upper- and
lower-triangular matrices
\be
\tilde L =\tilde L^{-1}_{+} \tilde L_{-}
\label{2.7}
\ee
where as before $\tilde l_{+} \tilde l_{-} =1$.

\noindent The matrices $L_{\pm}$ and $\tilde L_{\pm}$ are related to each other
by means of the following equations
\be
U^{-1}L_{\pm} =\tilde L^{-1}_{\pm} \tilde U
\label{2.8}
\ee
One can easily check that the matrices $\tilde L_{\pm} ,\tilde U$ have the
Poisson brackets (\ref{2.2a}-\ref{2.2c}) and we shall need the Poisson brackets
of $L_{\pm} , U$ and $\tilde L_{\pm} ,\tilde U$
\bea
\{L_{\a},L_{\b}\} &=&0 \qquad for \quad any \quad \a ,\b =+,-  \nonumber \\
\{\mc ,\tb\} &=&-\g\mc \tb \rb \nonumber \\
\{\Lc ,\tilde\tb\} &=&-\g\Lc \tilde\tb \rb \nonumber \\
\{\ta ,\tilde\tb\} &=&0
\label{3.9}
\eea

Up to now we considered \h of a complex Lie group.  For physical applications
one should single out some real form. If $\g$ is  imaginary ($\g^{*}=-\g$) one
can single out the
$SU(N)$ form by means of the standard anti-involution
\be
U^{*}=U^{-1},\quad L_{+}^{*}=L_{-}^{-1},\quad L_{-}^{*}=L_{+}^{-1}
\label{2.10a}
\ee
It is well-known that for real $\g$ the $SU(N)$ anti-involution $U^{*}=U^{-1}$
is not compatible with the Poisson structure (\ref{2.2a}). However
as was pointed out in ref.\cite{ms,af} the matrix $\tilde U$ can be used to
define an anti-involution on \h . Namely, taking into account that $\tilde U
\to
U^{-1}$ in the limit $\g \to 0$ one defines the anti-involution as follows
\be
U^{*}=\tilde U,\quad  L_{+}^{*}=L_{-},\quad L_{-}^{*}=L_{+}
\label{2.10}
\ee
The Heisenberg double of the $SL(N)$ group with this anti-involution reduces in
the limit $\g\to 0$ to $T^{*}SU(N)$. It can be easily checked that this
anti-involution is compatible with the Poisson structures
(\ref{2.2a}-\ref{2.2c}) and (\ref{3.9}). Let us notice that for real $\g$ the
involution which singles out the $SL(N,R)$ form looks as usual
\be
U^{*}= U,\quad L_{+}^{*}=L_{+},\quad L_{-}^{*}=L_{-}
\label{2.11}
\ee

Now we are ready to discuss a lattice gauge theory based on \h . So we place on
every link $({\bf n},{\bf e}_{i})$ a field taking values in \h $D_{+}^{\g}$.
Thus the phase space of the model is the direct product of $D_{+}^{\g}$ over
all
links: $\prod_{(\n, {\bf e}_{i})} D_{+}^{\g}(\n,i)$. We suppose that the
parameter $\g$ goes to zero in the continuum limit $a \to 0$. To define a
lattice gauge theory one should find lattice \G and a lattice Hamiltonian which
coincide with the \G and the Hamiltonian of the continuous theory in the
continuum limit. As was mentioned above the cotangent bundle $T^{*}G$ may be
considered as a limiting case of \h when $\g \to 0$. By this reason one could
look for such lattice
Gauss-law constraints which can be reduced to the lattice constraints
(\ref{13})
in the limit $\g \to 0$, $a$ being constant. The form of the constraints
depends
on the space dimension and, so we begin with the simplest case of
(1+1)-dimensional \Y theory on a cylinder.

In this case the constraints (2.13) look as follows
\be
G(n)=E(n)-\tilde E(n-1) =0
\label{3.12}
\ee
One could try to generalize the constraints $G(n)$ by replacing $E(n)$ on
$L(n)$
and $\tilde E(n)$ on $\tilde L(n)$. This replacement does work in the classical
theory, i.e. the corresponding constraints $G(n)=L(n)-\tilde L(n-1)$ are
first-class constraints. However one can show that quantization violates this
property of the constraints. To get a proper modification it seems to be
necessary to decompose $G(n)$ as follows
\be
G(n)=G_{+}(n)-G_{-}(n)
\label{3.13}
\ee
where $G_{\pm}(n)=E_{\pm}(n)-\tilde E_{\pm}(n)$ are upper- and lower-triangular
matrices and their diagonal parts $g_{\pm}(n)$ satisfy the following equation:
$g_{+}(n)+g_{-}(n)=0$.

\noindent Then the required modification (which seems to be the only possible)
is of the form
\be
G_{\pm}(n)=\tilde L_{\pm}(n-1)L_{\pm}(n)=1
\label{3.14}
\ee
One can easily calculate the Poisson brackets of the constraints $G_{\pm}(n)$
\bea
&&\{\ga ,\gb \} =\g [\rb ,\ga\gb ] \delta_{nm} \nonumber\\
&&\{\gaa ,\gbb \} =\g [\rb ,\gaa\gbb] \delta_{nm} \nonumber\\
&&\{\ga ,\gbb \} =\g [\r ,\ga\gbb]\delta_{nm}
\label{3.15}
\eea
We see that these Poisson brackets vanish on the constraints surface
$G_{\pm}(n)=1$ and therefore these constraints are first-class constraints. A
remarkable feature of these
constraints is that they form not a Lie-Poisson algebra but the same quadratic
Poisson algebra as the matrices $L_{\pm}$ do. In the limit $\g \to 0$,
$L_{\pm}\to 1+\g E_{\pm}$, $\tilde L_{\pm}\to 1-\g \tilde E_{\pm}$ one recovers
the old Gauss-law constraints (\ref{3.12}).

To complete the construction of the lattice theory one should find a lattice
Hamiltonian. As is well-known from the theory of Poisson-Lie groups
\cite{d1,s1,s2} generators of the ring of the Casimir functions of the Poisson
algebra (\ref{2.2b}) have the following form
\be
h_{k}=\tr L^{k}
\label{3.16}
\ee
It is clear that these functions are invariant with respect to the gauge
transformations which are generated by the constraints (\ref{3.14}).
In principle one can choose any combination of these functions as a Hamiltonian
of the theory
\be
H=\frac {a}{\g^{2}}\sum_{n=1}^{N}\sum_{k=-\infty}^{\infty}\, c_{k}(\tr
(L^{k}(n)-1))
\label{3.17}
\ee
Then in the limit $\g \to 0$ one gets the Hamiltonian
\be
H=a\sum_{n=1}^{N}\, \tr E^{2}(n)\sum_{k=-\infty}^{\infty}\, \frac {1}{2}
c_{k}k^{2}
\label{3.18}
\ee
which coincides up to a constant with the Hamiltonian used in the previous
section.

However there is a special choice of the Hamiltonian which leads on the quantum
level to a natural generalization of dynamics of the symmetric top
\be
H\sim \tr (\log L)^{2}
\label{3.19}
\ee
This Hamiltonian was implicitly used in ref.\cite{af} .

The functions (\ref{3.16}) are not the only ones gauge-invariant. Just as in
the
case of the usual lattice gauge theory one can construct Wilson line
observables. For the (1+1)-dimensional model the simplest observables look as
follows
\bea
W_{k_{1}\cdots k_{N}}&=&\tr (L^{k_{1}}(1)U(1)L^{k_{2}}(2)U(2)\cdots
L^{k_{N}}(N)U(N))  \nonumber \\
W^{\prime}_{k_{1}\cdots k_{N}}&=&\tr
(U^{-1}(N)L^{k_{N}}(N)U^{-1}(N-1)L^{k_{N-1}}(N-1)\cdots U^{-1}(1)L^{k_{1}}(1))
\label{2.20}
\eea
where $L=L_{+}L_{-}^{-1}$ and $k_{1},\cdots ,k_{N}$ are integers.

\noindent It is not difficult to show that the involutions
(\ref{2.10a}-\ref{2.11}) are compatible with the \G  (\ref{3.14}).

Thus we have constructed the (1+1)-dimensional gauge-invariant lattice \Y
theory
based on the assignment of \h to every link and now we pass to the
(2+1)-dimensional case. The usual
lattice Gauss-law constraints (\ref{13}) look in this case as follows
\be
G_{\pm}(n_{1},n_{2})=E_{\pm}(n_{1},n_{2};1)+E_{\pm}(n_{1},n_{2};2)-\tilde
E_{\pm}(n_{1}-1,n_{2};1)-\tilde E_{\pm}(n_{1},n_{2}-1;2)
\label{2.21}
\ee
where just as for the two-dimensional case we decompose $G$ into upper- and
lower-triangular matrices.

There are (at least) six nonequivalent modifications of the constraints
(\ref{2.21}). In this paper we use the following constraints
\be
G_{\pm}(n_{1},n_{2})=\tilde L_{\pm}(n_{1}-1,n_{2};1)\tilde
L_{\pm}(n_{1},n_{2}-1;2)L_{\pm}(n_{1},n_{2};1)L_{\pm}(n_{1},n_{2};2)=1
\label{2.22}
\ee
These constraints form the same Poisson algebra as for the two-dimensional case
with the only change $n \to \n =(n_{1},n_{2})$ in eq.(\ref{3.15}). Let us
notice
that the product of matrices $L_{\pm}$, $\tilde L_{\pm}$ is taken in the
clock-wise order (starting from $\tilde L_{\pm}(n_{1}-1,n_{2};1)$). Namely due
to this choice of the constraints (\ref{2.22})  the following expressions
generalizing the Wilson term (\ref{10})
\bea
&&W(n_{1}+1,n_{2})= \nonumber \\
&& \tr (G^{-1}_{\pm}(n_{1}+1,n_{2})\tilde U(n_{1},n_{2};1)
U(n_{1},n_{2};2)U(n_{1},n_{2}+1;1)\tilde U(n_{1}+1,n_{2};2))  \nonumber \\
&&W^{\prime}(n_{1}+1,n_{2}) =  \nonumber \\
&& \tr (G_{\pm}(n_{1}+1,n_{2})\tilde U^{-1}(n_{1}+1,n_{2};2)
U^{-1}(n_{1},n_{2}+1;1)U^{-1}(n_{1},n_{2};2)\tilde U^{-1}(n_{1},n_{2};2))
\nonumber \\
\label{2.23}
\eea
are gauge-invariant.

\noindent Let us mention that one can find a similar generalization of an
arbitrary Wilson line observable but the corresponding formula depends on the
local form of the Wilson line and will not be discussed in the paper.

These Wilson terms can be used to write down a gauge-invariant Hamiltonian
\be
H=H(L)+\frac {1}{2a^{2}e^{2}}\tr\sum_{n_{1},n_{2}} (W(n_{1},n_{2}) +W^{\prime}
(n_{1},n_{2}))
\label{2.24}
\ee
The first term in eq.(\ref{2.24}) depends only on $L_{\pm}$ and coincides with
$\sum_{\n ,i}\tr E^{2}(\n ,i)$ in the limit $\g \to 0$ (see the discussion of
the two-dimensional model).

This completes the construction of the (2+1)-dimensional lattice \Y theory. For
imaginary $\g$ one can impose the anti-involution (\ref{2.10a}) to get the real
form $SU(N)$. However for real $\g$ the anti-involution (\ref{2.10}) is not
compatible with the \G and thus we can get only the real form $SL(N,R)$. Let us
notice that in these cases the Hamiltonian is real.

\section{Discussion}

In this paper we considered the gauge-invariant Hamiltonian formulation of
classical lattice \Y theory in (1+1) and (2+1) dimensions. In this formulation
we placed on every link
\h of a Lie group. It is clear that the construction presented can be
generalized to the case of the (d+1)-dimensional $SL(N)$ \Y model \cite{fr} .

For the (1+1)-dimensional model it appeares to be possible to make the limit
$a\to 0$, $\g$ being constant. The theory obtained in such a way seems to be
related to Poisson-$\sigma$-models recently introduced in ref.\cite{ss} . It
would be interesting to clarify this relation.

We have not written down the Lagrangians which correspond to these lattice
gauge
models. In principle one can do it by using some expressions for the symplectic
form of \h which were found in ref.\cite{am}.

We considered only classical theory and it would be of great interest to
quantize the models. There is no problem in quantizing the Poisson structure of
the Heisenberg double. One just gets the quantized algebra of functions on \h
which was introduced in ref.\cite{af2} . The classical $r$-matrices $\rb$ are
to
be replaced by the $R$-matrices $R_{\pm}(q)=1+i\hbar\g\rb +\cdots$, where
$q={\rm e}^{i\hbar\g}$. Thus a real $\g$ corresponds to $q$ lying on the unit
circle of the complex plane and an imaginary $\g$ corresponds to a real $q$. It
is not difficult to verify that in quantum theory the \G [3.38] and [3.46] are
first-class constraints and commute with the quantum Hamiltonians [3.41] and
[3.48] if these Hamiltonians are finite polynomials of $L$ and $L^{-1}$. It
seems that the case of $q$ being a root of unity, $q^{N}=1$, is the most
interesting one. However the present formulation permits to single out only
$SL(N,R)$ real form for such a $q$. It is an interesting problem to find a
similar formulation for the $SU(N)$ case. Let us finally notice that
$q$-deformed lattice gauge theory was considered in refs.\cite{fock,b,ags,br}
in
connection with the Chern-Simons theory.

Due to the fact that there are two arbitrary parameters $a$ and $\g$ in the
theory one may expect that the theory has more rich phase structure than the
usual lattice gauge models. We hope to consider these problems in forthcoming
publications.

{\bf Acknowledgements:} The author would like to thank G.Fiore, P.Schupp and
\break A.A.Slavnov for discussions. He is grateful to
Professor J.Wess for kind hospitality and the Alexander von Humboldt
Foundation for the support. This work has been supported in part by
ISF-grant MNB000 and by the Russian Basic Research Fund under
grant number 94-01-00300a.

\end{document}